\documentstyle[twoside,epsf]{article}

\catcode`\@=11
\long\def\@makefntext#1{
\protect\noindent \hbox to 3.2pt {\hskip-.9pt  
$^{{\eightrm\@thefnmark}}$\hfil}#1\hfill}		

\def\thefootnote{\fnsymbol{footnote}}
\def\@makefnmark{\hbox to 0pt{$^{\@thefnmark}$\hss}}	
	
\def\ps@myheadings{\let\@mkboth\@gobbletwo
\def\@oddhead{\hbox{}
\rightmark\hfil\eightrm\thepage}   
\def\@oddfoot{}\def\@evenhead{\eightrm\thepage\hfil
\leftmark\hbox{}}\def\@evenfoot{}
\def\sectionmark##1{}\def\subsectionmark##1{}}

\oddsidemargin=\evensidemargin
\renewcommand{\thefootnote}{\fnsymbol{footnote}}

\newcounter{sectionc}\newcounter{subsectionc}\newcounter{subsubsectionc}
\renewcommand{\section}[1] {\vspace{12pt}\addtocounter{sectionc}{1} 
\setcounter{subsectionc}{0}\setcounter{subsubsectionc}{0}\noindent 
	{\tenbf\thesectionc. #1}\par\vspace{5pt}}
\renewcommand{\subsection}[1] {\vspace{12pt}\addtocounter{subsectionc}{1} 
	\setcounter{subsubsectionc}{0}\noindent 
	{\bf\thesectionc.\thesubsectionc. {\kern1pt \bfit #1}}\par\vspace{5pt}}
\renewcommand{\subsubsection}[1] {\vspace{12pt}\addtocounter{subsubsectionc}{1}
	\noindent{\tenrm\thesectionc.\thesubsectionc.\thesubsubsectionc.
	{\kern1pt \tenit #1}}\par\vspace{5pt}}
\newcommand{\nonumsection}[1] {\vspace{12pt}\noindent{\tenbf #1}
	\par\vspace{5pt}}

\newcounter{appendixc}
\newcounter{subappendixc}[appendixc]
\newcounter{subsubappendixc}[subappendixc]
\renewcommand{\thesubappendixc}{\Alph{appendixc}.\arabic{subappendixc}}
\renewcommand{\thesubsubappendixc}
	{\Alph{appendixc}.\arabic{subappendixc}.\arabic{subsubappendixc}}

\renewcommand{\appendix}[1] {\vspace{12pt}
        \refstepcounter{appendixc}
        \setcounter{figure}{0}
        \setcounter{table}{0}
        \setcounter{lemma}{0}
        \setcounter{theorem}{0}
        \setcounter{corollary}{0}
        \setcounter{definition}{0}
        \setcounter{equation}{0}
        \renewcommand{\thefigure}{\Alph{appendixc}.\arabic{figure}}
        \renewcommand{\thetable}{\Alph{appendixc}.\arabic{table}}
        \renewcommand{\theappendixc}{\Alph{appendixc}}
        \renewcommand{\thelemma}{\Alph{appendixc}.\arabic{lemma}}
        \renewcommand{\thetheorem}{\Alph{appendixc}.\arabic{theorem}}
        \renewcommand{\thedefinition}{\Alph{appendixc}.\arabic{definition}}
        \renewcommand{\thecorollary}{\Alph{appendixc}.\arabic{corollary}}
        \renewcommand{\theequation}{\Alph{appendixc}.\arabic{equation}}
        \noindent{\tenbf Appendix \theappendixc #1}\par\vspace{5pt}}
\newcommand{\subappendix}[1] {\vspace{12pt}
        \refstepcounter{subappendixc}
        \noindent{\bf Appendix \thesubappendixc. {\kern1pt \bfit #1}}
	\par\vspace{5pt}}
\newcommand{\subsubappendix}[1] {\vspace{12pt}
        \refstepcounter{subsubappendixc}
        \noindent{\rm Appendix \thesubsubappendixc. {\kern1pt \tenit #1}}
	\par\vspace{5pt}}

\newcommand{\textlineskip}{\baselineskip=13pt}
\newcommand{\smalllineskip}{\baselineskip=10pt}

\def\eightcirc{
\begin{picture}(0,0)
\put(4.4,1.8){\circle{6.5}}
\end{picture}}
\def\eightcopyright{\eightcirc\kern2.7pt\hbox{\eightrm c}} 

\newcommand{\copyrightheading}[1]
	{\vspace*{-2.5cm}\smalllineskip{\flushleft
	{\footnotesize International Journal of Modern Physics C, #1}\\
	{\footnotesize $\eightcopyright$\, World Scientific Publishing
	 Company}\\
	 }}

\newcommand{\publisher}[2]{{\begin{center}\footnotesize\smalllineskip 
	Received #1\\
	Revised #2
	\end{center}
	}}

\def\abstracts#1#2#3{{
	\centering{\begin{minipage}{4.5in}\baselineskip=10pt\footnotesize
	\parindent=0pt #1\par 
	\parindent=15pt #2\par
	\parindent=15pt #3
	\end{minipage}}\par}} 

\def\keywords#1{{
	\centering{\begin{minipage}{4.5in}\baselineskip=10pt\footnotesize
	{\footnotesize\it Keywords}\/: #1
	\end{minipage}}\par}}

\renewenvironment{thebibliography}[1]
        {\frenchspacing
	 \ninerm\baselineskip=11pt
         \begin{list}{\arabic{enumi}.}
        {\usecounter{enumi}\setlength{\parsep}{0pt}     
	 \setlength{\leftmargin 12.7pt}{\rightmargin 0pt} 
         \setlength{\itemsep}{0pt} \settowidth
	{\labelwidth}{#1.}\sloppy}}{\end{list}}

\newcounter{itemlistc}
\newcounter{romanlistc}
\newcounter{alphlistc}
\newcounter{arabiclistc}

\newcommand{\fcaption}[1]{
        \refstepcounter{figure}
        \setbox\@tempboxa = \hbox{\footnotesize Fig.~\thefigure. #1}
        \ifdim \wd\@tempboxa > 5in
           {\begin{center}
        \parbox{5in}{\footnotesize\smalllineskip Fig.~\thefigure. #1}
            \end{center}}
        \else
             {\begin{center}
             {\footnotesize Fig.~\thefigure. #1}
              \end{center}}
        \fi}

\newcommand{\tcaption}[1]{
        \refstepcounter{table}
        \setbox\@tempboxa = \hbox{\footnotesize Table~\thetable. #1}
        \ifdim \wd\@tempboxa > 5in
           {\begin{center}
        \parbox{5in}{\footnotesize\smalllineskip Table~\thetable. #1}
            \end{center}}
        \else
             {\begin{center}
             {\footnotesize Table~\thetable. #1}
              \end{center}}
        \fi}

\def\@citex[#1]#2{\if@filesw\immediate\write\@auxout
	{\string\citation{#2}}\fi
\def\@citea{}\@cite{\@for\@citeb:=#2\do
	{\@citea\def\@citea{,}\@ifundefined
	{b@\@citeb}{{\bf ?}\@warning
	{Citation `\@citeb' on page \thepage \space undefined}}
	{\csname b@\@citeb\endcsname}}}{#1}}

\newif\if@cghi
\def\cite{\@cghitrue\@ifnextchar [{\@tempswatrue
	\@citex}{\@tempswafalse\@citex[]}}
\def\citelow{\@cghifalse\@ifnextchar [{\@tempswatrue
	\@citex}{\@tempswafalse\@citex[]}}
\def\@cite#1#2{{$\null^{#1}$\if@tempswa\typeout
	{IJCGA warning: optional citation argument 
	ignored: `#2'} \fi}}

\def\pmb#1{\setbox0=\hbox{#1}
	\kern-.025em\copy0\kern-\wd0
	\kern.05em\copy0\kern-\wd0
	\kern-.025em\raise.0433em\box0}

\def\fnm#1{$^{\mbox{\scriptsize #1}}$}
\def\fnt#1#2{\footnotetext{\kern-.3em
	{$^{\mbox{\scriptsize #1}}$}{#2}}}

\def\fpage#1{\begingroup
\voffset=.3in
\thispagestyle{empty}\begin{table}[b]\centerline{\footnotesize #1}
	\end{table}\endgroup}

\def\runninghead#1#2{\pagestyle{myheadings}
\markboth{{\protect\footnotesize\it{\quad #1}}\hfill}
{\hfill{\protect\footnotesize\it{#2\quad}}}}
\font\tenrm=cmr10
\font\tenit=cmti10 
\font\tenbf=cmbx10
\font\bfit=cmbxti10 at 10pt
\font\ninerm=cmr9

\font\eightrm=cmr8

\def\qed{\hbox{${\vcenter{\vbox{			
   \hrule height 0.4pt\hbox{\vrule width 0.4pt height 6pt
   \kern5pt\vrule width 0.4pt}\hrule height 0.4pt}}}$}}

\renewcommand{\thefootnote}{\fnsymbol{footnote}}	

\def\bsc{{\sc a\kern-6.4pt\sc a\kern-6.4pt\sc a}}  	
\def\bflatex{\bf L\kern-.30em\raise.3ex\hbox{\bsc}\kern-.14em 
T\kern-.1667em\lower.7ex\hbox{E}\kern-.125em X} 

\begin{document}

\runninghead{Percolation thresholds in high dimensions}
            {Percolation thresholds in high dimensions}
\normalsize\textlineskip
\thispagestyle{empty}
\setcounter{page}{1}
\copyrightheading{}	
\vspace*{0.88truein}

\fpage{1}
\centerline{\bf CALCULATION OF PERCOLATION THRESHOLDS IN HIGH DIMENSIONS}
\vspace*{0.035truein}
\centerline{\bf FOR FCC, BCC AND DIAMOND LATTICES}
\vspace*{0.37truein}
\centerline{\footnotesize Steven C. van der Marck}
\vspace*{0.015truein}
\centerline{\footnotesize\it SIEP Research and Technical Services, P.O. Box 60}
\baselineskip=10pt
\centerline{\footnotesize\it 2280 AB Rijswijk, The Netherlands.}
\baselineskip=10pt
\centerline{\footnotesize\it e-mail: s.c.vandermarck@siep.shell.com}
\vspace*{0.225truein}
\publisher{March 24, 1998}{}

\vspace*{0.21truein}
\abstracts{
           Site and bond percolation thresholds are calculated
           for the face centered cubic, body centered cubic and
           diamond lattices in 4, 5, and 6 dimensions.
           The results are used to study the behaviour of
           percolation thresholds as a functions of dimension.
           It is shown that the predictions from a recently
           proposed invariant for percolation thresholds
           are not satisfactory for these lattices.
          }{}{}

\vspace*{10pt}

\vspace*{1pt}\textlineskip	
\section{Introduction}		
\vspace*{-0.5pt}

\textheight=7.8truein
\setcounter{footnote}{0}
\renewcommand{\thefootnote}{\alph{footnote}}

\noindent
Percolation problems have a wide range of applicability,
and have therefore attracted a fair bit of attention
over many years\cite{stauffer}.
Nevertheless the percolation thresholds, which are
among the basic quantities for percolation on lattices,
have been calculated exactly for only 
a few two-dimensional lattices.
For many other lattices these thresholds have been
calculated numerically.
These numerical values can then be analysed, to determine
regular behaviour, or trends, as a function of the
lattice coordination number, dimensionality, etc.
A limited review of such efforts has been given
elsewhere\cite{galam_96}.

Recently, Galam and Mauger\cite{galam}
proposed an invariant for site and bond percolation thresholds,
$p_{cs}$ and $p_{cb}$ respectively.
The proposed invariant reads
\begin{equation}
   (p_{cs})^{1/a_s} (p_{cb})^{-1/a_b} = \frac{\delta}{d},
\end{equation}
where $d$ is the dimension of the lattice,
and $a_s$, $a_b$, and $\delta$ are positive constants.
The lattices studied by Galam and Mauger were divided into
two classes, and for each of these classes the values
of the parameters were fitted.
The values for the first class were
$ \{ a_s = 0.3670; a_b = 0.6897; \delta = 1.3638 \}$,
while for the second class
$ \{ a_s = 0.6068; a_b = 0.9346; \delta = 1.9340 \}$.
For the lattices used by Galam and Mauger,
the numerical results for this invariant are
indeed constant within~5\%.
Although the deviations up to 5\% cannot be explained
by the inaccuracy in the values for the percolation thresholds,
it is an interesting observation that the above combination
of percolation thresholds yields almost constant values.
Especially the absence of coordination number~$q$ in the invariant
makes it valuable.

However, the invariant can be tested on more lattices for which the
percolation thresholds are known.
We list data for many lattices in this article,
obtained from various sources in literature.
However, for some of the higher dimensional lattices, only site- or
only bond percolation thresholds were known.
Therefore the `missing' values are computed here, so that
the invariant can be calculated for these lattices as well.
Moreover, it is interesting to study the scaling
of percolation thresholds as a function of
dimension for some of the important lattices, like
the face centered cubic lattice (fcc),
the body centered cubic lattice (bcc) and the diamond lattice.

This paper is organised as follows.
In section~2, definitions are given for the fcc,
bcc, and diamond lattices in higher dimensions.
The cluster algorithm and some implementation issues are
discussed in section~3, after which the results
are presented in section~4.
Finally, section~5 contains a discussion of the results.


\section{Lattice definitions in higher dimensions}
\label{sec_lattice}
\noindent
For the definitions of lattices in $d$~dimensions, we refer
to Conway and Sloane\cite{conway_97} for the
face centered cubic and body centered cubic lattice, and
to Van der Marck\cite{marck_98} for the diamond and Kagom\'e lattice.
For completeness, we briefly describe here the
generalisation of the fcc, bcc and diamond lattices to higher
dimensions.
First we need to have, for each lattice, a set of $d$
independent lattice vectors.
A point is called a lattice site if and only if it is an integer
combination of these vectors.
Secondly, we want to know the neighbours of a site, described
in terms of the lattice vectors.
We will denote an orthonormal basis of $d$-dimensional space
by ${\bf x}_i$.

\begin{table}[tb]
{\footnotesize
  \begin{center}
  \tcaption{The site percolation thresholds of $d$-dimensional {\bf fcc}
             lattices as a function of linear lattice size~$L$.
             The values in the last row are results of a fit 
             to the scaling relation
             $ | p_c(L) - p_c(\infty) | \propto L^{-1/\nu} $.
             Error estimates concerning the last digit are indicated
             between brackets.
            }
  \label{tab_fcc}
  \begin{tabular}{rcrcrcrc}
    \hline\noalign{\smallskip}
    \multicolumn{2}{c}{$d=3$} & \multicolumn{2}{c}{$d=4$} &
    \multicolumn{2}{c}{$d=5$} & \multicolumn{2}{c}{$d=6$} \\
    $L$ & $p_{cs}$ & $L$ & $p_{cs}$ &
    $L$ & $p_{cs}$ & $L$ & $p_{cs}$ \\
    \hline\noalign{\smallskip}
      16 & 0.2097(2) & 16 & 0.0856(2) &  8 & 0.0457(2) &  8 & 0.0257(2) \\
      32 & 0.2035(2) & 24 & 0.0847(2) & 12 & 0.0435(2) & 10 & 0.0258(2) \\
      64 & 0.2016(2) & 32 & 0.0845(2) & 16 & 0.0432(2) & 12 & 0.0252(2) \\
     128 & 0.2001(2) & 48 & 0.0844(2) & 24 & 0.0431(2) & 14 & 0.0252(2) \\
     250 & 0.1998(2) & 64 & 0.0843(2) & 32 & 0.0432(2) & 16 & 0.0252(2) \\
    \noalign{\smallskip}
    $\infty$ & 0.1994(2) & $\infty$ & 0.0842(3) &
    $\infty$ & 0.0431(3) & $\infty$ & 0.0252(5) \\
    \hline\noalign{\smallskip}
  \end{tabular}
  \end{center}
}
\end{table}
The $d$-dimensional fcc lattice is the set of points
in $Z\!\!\!Z^n$, for which the sum of the coordinates is even.
Conway and Sloane use the notation `$D_d$' for this
lattice\cite{conway_97}.
(Note that $D_d$ is the `closest packed' lattice
for $d=3$, $4$ and $5$, but not for higher
dimensions\cite{conway_95}.)
Each site has $2d(d-1)$  neighbours at a relative location
$ \pm {\bf x}_i \pm {\bf x}_j $, for $i, j = 1,\ldots,d$ ($i \neq j$).
As a set of lattice vectors one can choose
${\bf f}_i = {\bf x}_1 + {\bf x}_i$.
The neighbours of a site are given in terms of these
lattice vectors as
\begin{equation}
  \left.
    \begin{array}{l}
      \pm ( {\bf f}_i - {\bf f}_j ) \\
      \pm ( {\bf f}_i + {\bf f}_j - {\bf f}_1 )
    \end{array}
  \right\}
  \mbox{ for all  } i, j = 1,\ldots,d \hspace*{5mm}(i \neq j).
  \label{eq_nb_fcc}
\end{equation}

A site in the $d$-dimensional bcc lattice has $2^d$ neighbours,
located at
$ \frac{1}{2} ( \pm {\bf x}_1 \pm {\bf x}_2 \pm \ldots
                \pm {\bf x}_d ) $.
This is dubbed the `generalized bcc net' by Conway and
Sloane\cite{conway_97}.
A possible set of lattice vectors is
$ {\bf b}_i = - \frac{1}{2}( {\bf x}_1 + \ldots + {\bf x}_{i-1} )
              + \frac{1}{2}( {\bf x}_i + \ldots + {\bf x}_{d} ) $
for $i=1,\ldots,d$.
The neighbours of a site are given in terms of these vectors by
\begin{equation}
    \begin{array}{ll}
      \pm   {\bf b}_i               & \hspace*{5mm}\mbox{for } i=1,\ldots,d,\\
      \pm ( {\bf b}_i - {\bf b}_j + {\bf b}_k ) &
                                      \hspace*{5mm}\mbox{for } i>j>k, \\
      \pm ( {\bf b}_i - {\bf b}_j + {\bf b}_k
          - {\bf b}_l + {\bf b}_m ) & \hspace*{5mm}\mbox{for } i>j>k>l>m, \\
      \ldots
    \end{array}
\end{equation}
\begin{table}[tb]
{\footnotesize
  \begin{center}
  \tcaption{The site percolation thresholds of $d$-dimensional {\bf bcc}
             lattices as a function of linear lattice size~$L$.
             The values in the last row are results of a fit 
             to the scaling relation
             $ | p_c(L) - p_c(\infty) | \propto L^{-1/\nu} $.
             Error estimates concerning the last digit are indicated
             between brackets.
            }
  \label{tab_bcc}
  \begin{tabular}{rcrcrcrc}
    \hline\noalign{\smallskip}
    \multicolumn{2}{c}{$d=3$} & \multicolumn{2}{c}{$d=4$} &
    \multicolumn{2}{c}{$d=5$} & \multicolumn{2}{c}{$d=6$} \\
    $L$ & $p_{cs}$ & $L$ & $p_{cs}$ &
    $L$ & $p_{cs}$ & $L$ & $p_{cs}$ \\
    \hline\noalign{\smallskip}
      16 & 0.2593(2) & 16 & 0.1078(2) &  8 & 0.0497(2) &  8 & 0.0215(2) \\
      32 & 0.2514(2) & 24 & 0.1058(2) & 12 & 0.0468(2) & 10 & 0.0205(2) \\
      64 & 0.2483(2) & 32 & 0.1050(2) & 16 & 0.0459(2) & 12 & 0.0203(2) \\
     128 & 0.2471(2) & 48 & 0.1043(2) & 24 & 0.0453(2) & 14 & 0.0201(2) \\
     250 & 0.2463(2) & 64 & 0.1042(2) & 32 & 0.0450(2) & 16 & 0.0202(2) \\
    \noalign{\smallskip}
    $\infty$ & 0.2458(2) & $\infty$ & 0.1037(3) &
    $\infty$ & 0.0446(4) & $\infty$ & 0.0199(5) \\
    \hline\noalign{\smallskip}
  \end{tabular}
  \end{center}
}
\end{table}

The $d$-dimensional diamond lattice is a lattice with a 2-point basis.
Let us call the two points in the basis A and~B.
The full lattice is built by translation of the lattice basis
over $d$ independent vectors~${\bf t}_i$.
Each A-site has $d+1$ neighbours of type~B.
One of these neighbours is the B-site in the same basis,
while the other $d$ neighbours are the B-type sites at a
relative location ${\bf t}_i$.
Each B-site also has $d+1$ neighbours, one of which is the
A-site in the same basis and the others are the A-type 
sites at a relative location $-{\bf t}_i$.



\section{Lattice coding and cluster algorithm}
\label{sec_algor}
\noindent
Two programs to calculate percolation thresholds were developed.
One program was geared towards the handling of any desired
lattice topology, the other towards speed and efficient memory
usage for certain specific
lattices\fnm{a}\fnt{a}{The programs, one in Fortran and one in C,
can be obtained from the author.}.

In the general purpose program to calculate percolation thresholds
we used a generic method to specify a lattice.
The bonds connected to lattice sites were coded in Fortran as
{\tt NB\_SITES(QMAX,NSITES)}, where {\tt NSITES} is the total
number of sites in the lattice, and {\tt QMAX} is the maximum
number of bonds connected to a site (maximum coordination number).
The integer value of {\tt NB\_SITES(I,S)} was set to the
bond number of the $i^{th}$ bond connected to site~$s$.

The sites connected to bonds were coded as
{\tt NB\_BONDS(2,NBONDS)}, where
{\tt NBONDS} is the total number of bonds in the lattice.
In other words, link~$i$ connects site {\tt NB\_BONDS(1,I)}
and {\tt NB\_BONDS(2,I)}.
The two arrays {\tt NB\_SITES} and {\tt NB\_BONDS} together specify
the topology of the lattice completely.

When we want to calculate a percolation threshold for
such a lattice, we need a cluster algorithm, e.g. the
one proposed by Hoshen and Kopelman\cite{hoshen}.
This algorithm works by assigning
a cluster label~$m$ to each site and bond.
(It is assumed that $m>0$.)
In addition to the arrays to hold these labels,
the algorithm uses one other array~$N(m)$.
For each proper label, $N(m)$ is greater than zero
(by definition of `proper'), in which case it holds the number
of sites and bonds that belong to the cluster~$m$.
For a non-proper label $N(m)$ is negative,
and $-N(m)$ refers to a label~$m^{\prime}$,
which is the cluster that cluster~$m$ has been merged with.

\begin{table}[tb]
{\footnotesize
  \begin{center}
  \tcaption{The site percolation thresholds of $d$-dimensional {\bf diamond}
             lattices as a function of linear lattice size~$L$.
             The values in the last row are results of a fit 
             to the scaling relation
             $ | p_c(L) - p_c(\infty) | \propto L^{-1/\nu} $.
             Error estimates concerning the last digit are indicated
             between brackets.}
  \label{tab_dia}
  \begin{tabular}{rcrcrcrcrc}
    \hline\noalign{\smallskip}
    \multicolumn{2}{c}{$d=2$} & \multicolumn{2}{c}{$d=3$} &
    \multicolumn{2}{c}{$d=4$} & \multicolumn{2}{c}{$d=5$} &
    \multicolumn{2}{c}{$d=6$} \\
    $L$ & $p_{cs}$ & $L$ & $p_{cs}$ & $L$ & $p_{cs}$ &
    $L$ & $p_{cs}$ & $L$ & $p_{cs}$ \\
    \hline\noalign{\smallskip}
 256& 0.6964(2)&  16& 0.4398(2)& 16& 0.3014(2)&  8& 0.2268(2)&  8& 0.1760(2)\\
 512& 0.6964(2)&  32& 0.4347(2)& 24& 0.2996(2)& 12& 0.2251(2)& 10& 0.1762(2)\\
1024& 0.6969(2)&  64& 0.4317(2)& 32& 0.2989(2)& 16& 0.2248(2)& 12& 0.1770(2)\\
2048& 0.6970(2)& 128& 0.4312(2)& 50& 0.2984(2)& 24& 0.2249(2)& 14& 0.1777(2)\\
3072& 0.6970(1)& 250& 0.4306(2)& 64& 0.2983(2)& 30& 0.2251(2)& 16& 0.1783(2)\\
    \noalign{\smallskip}
    $\infty$ & 0.6971(2) & $\infty$ & 0.4301(2) &
    $\infty$ & 0.2978(2) & $\infty$ & 0.2252(3) & $\infty$ & 0.1799(5) \\
    \hline\noalign{\smallskip}
  \end{tabular}
  \end{center}
}
\end{table}
The algorithm consists of the following steps.
\begin{itemize}
  \item Initialise by setting $N(m) \leftarrow 0$ for all $m$,
        and $m_{next} \leftarrow 1$.
  \item Loop over all sites and bonds of the lattice.
  \item For each new occupied site and bond one encounters,
        one has to determine
        which cluster it belongs to. To this end, loop over the
        neighbours that one already determined a cluster label for.
        Three different situations can occur.
        \begin{description} 
          \item[=0] If there are no labeled neighbours,
                    one has to define a new cluster label.
                    Use $m \leftarrow m_{next}$, set
                    $N(m)     \leftarrow 1$, and
                    $m_{next} \leftarrow m_{next}+1$.
          \item[=1] If there is one labeled cluster~$m$, the current
                    site/bond will get the label of that cluster.
                    Set $N(m) \leftarrow N(m) + 1$.
          \item[$\geq$ 2] If there are two or more neighbours that belong
                    to different clusters, then those clusters have to
                    be `merged':
                    \begin{itemize}
                      \item Determine the lowest cluster label, $m_{low}$,
                            of these neighbours.
                      \item For all other labels $m_{other}$, set
                            $N(m_{other}) \leftarrow - m_{low}$, and \\
                            $N(m_{low}) \leftarrow N(m_{low})+N(m_{other})$.
                      \item For the current site or bond, set
                            $m \leftarrow m_{low}$, and
                            $N(m) \leftarrow N(m) + 1$.
                    \end{itemize}
        \end{description} 
\end{itemize}
In this algorithm, it is important to know the neighbours of a site or bond.
{\it The specification of neighbours
     is the only item that is lattice-specific.}
When one is interested in site-bond percolation,
the neighbours of a site are directly given by
the {\tt NB\_SITES} array, and the neighbours of a bond
by the {\tt NB\_BONDS} array.
Therefore the implementation of the Hoshen-Kopelman algorithm
is fairly straightforward for site-bond problems.
There is but one important difference with the algorithm
as described by Hoshen and Kopelman, which is that in their
article, the networks could be traversed in an ordered way,
whereas here we cannot.
Let us take as an example the square lattice in two dimensions.
If the loop over sites is performed as
\begin{verbatim}
      DO I = 1, L
         DO J = 1 , L
            ...
         ENDDO
      ENDDO
\end{verbatim}
it is clear that neighbouring sites $\{i,j-1\}$ and
$\{i-1,j\}$ have been visited before site $\{i,j\}$,
whereas $\{i+1,j\}$ and $\{i,j+1\}$ have not.

Here, on the other hand,
we are dealing with a generic lattice, which has no
definite order, and hence one cannot traverse the lattice
from one side to the other in a regular way.
Therefore it is not a priori clear which neighbours
have already been visited and hence have been assigned
a cluster label.
This problem can be solved by a suitable initialisation.
For instance, we can define arrays to hold the cluster label
per site ({\tt LS(NSITES)}) and per bond ({\tt LB(NBONDS)}).
During initialisation we can set {\tt LS(I)=0} for
occupied sites and {\tt LS(I)=-1} for empty ones,
and similarly for bonds.
In the couse of the algorithm, the sites for which
{\tt LS(I)=0} will be assigned a valid label
{\tt LS(I)=}$m$.
Since valid labels are greater than zero, one can,
at any time during the algorithm, recognise which
sites are `empty', which ones are `occupied but unassigned',
and which ones are `assigned'.

Having performed the cluster algorithm, one still has to
decide whether there is a percolating cluster or not.
One has to define, therefore, what is percolation
on this lattice: from where to where does a cluster have
to extend, to be called percolating?
In other words, one should define a beginning of the lattice
and an end (or an `IN' and an `OUT').
For example, one can arrange the list of sites
such that the last $N_{out}$
sites will be the exit of the lattice and the last $N_{in}$
before that will be the entrance.
Having done so, one can loop over the entrance sites
and mark the clusters that are connected to it.
The final step is then to
loop over the exit sites, to see whether there is
a `marked' cluster connected to the exit.
If so, there is a percolating cluster, and otherwise there
is not.

The marking of a cluster can be done in various ways,
for example by adding $N_{s+b}$ to $N(m)$, where
$N_{s+b}$ is the total number of bonds and sites.
Because the total number of sites and bonds that can
belong to any cluster is at maximum $N_{s+b}$,
this method is allowed:
any cluster with $N(m) > N_{s+b}$ is marked as connected
to the entrance, and consists of $N(m)-N_{s+b}$ sites and bonds.
\begin{table}[tb]
{\footnotesize
  \begin{center}
  \tcaption{The {\it bond} percolation thresholds of
             $d$-dimensional Kagom\'e, fcc and bcc
             lattices as a function of linear lattice size~$L$.
             The values in the row marked with $\infty$
             are results of a fit to the scaling relation
             $ | p_c(L) - p_c(\infty) | \propto L^{-1/\nu} $.
             Error estimates concerning the last digit are indicated
             between brackets.
             The values in the last row were obtained with an
             alternative method for extrapolation to $L=\infty$
             (see text).
            }
  \label{tab_bond}
  \begin{tabular}{rcccrccc}
    \hline\noalign{\smallskip}
    \multicolumn{4}{c}{$d=4$} &
    \multicolumn{4}{c}{$d=5$} \\
    $L$ & Kagom\'e & fcc & bcc & $L$ & Kagom\'e & fcc & bcc \\
    \hline\noalign{\smallskip}
       8 & 0.1725(2) & 0.0473(2) & 0.0729(2) & \hphantom{xxx}
       6 & 0.1176(2) & 0.0227(2) & 0.0287(2) \\
      12 & 0.1755(2) & 0.0484(2) & 0.0738(2) &
       8 & 0.1237(2) & 0.0237(2) & 0.0301(2) \\
      16 & 0.1764(2) & 0.0487(2) & 0.0738(2) &
      10 & 0.1264(2) & 0.0252(2) & 0.0309(2) \\
      24 & 0.1767(2) & 0.0490(2) & 0.0740(2) &
      12 & 0.1275(2) & 0.0255(2) & 0.0314(2) \\
    \noalign{\smallskip}
    $\infty$ & 0.177(1) & 0.049(1) & 0.074(1) &
    $\infty$ & 0.130(2) & 0.026(2) & 0.033(1) \\
    \noalign{\smallskip}
    check & 0.179(2) & 0.049(1) & 0.075(1) &
    check & 0.132(2) & 0.027(1) & 0.033(1) \\
    \hline\noalign{\smallskip}
  \end{tabular}
  \end{center}
}
\end{table}

The cases of site-site percolation or bond-bond percolation are
special cases of site-bond percolation.
One can e.g. calculate the bond-bond percolation thresholds by
simply making sure that all sites are `occupied'.

The special purpose program to calculate the site percolation
thresholds for several lattices is both faster and more
economic in the use of memory, and was written in~C.
For the Bravais lattices,
this is achieved by only using an array for the sites
{\tt sites(x,y,$\ldots$)} and an array for~$N(m)$.
These lattices can be traversed in a structured way,
so that we know in advance which neighbours have been
visited before, and which ones haven't.
The specification of the neighbours of a site is done,
for the four-dimensional fcc lattice, by
\begin{verbatim}
   int ixp[] = {  1, 1, 0, 1, 0, 0,   0, 0, 0, 1, 1, 1 };
   int iyp[] = { -1, 0, 1, 0, 1, 0,  -1, 0, 0,-1,-1, 0 };
   int izp[] = {  0,-1,-1, 0, 0, 1,   0,-1, 0,-1, 0,-1 };
   int iap[] = {  0, 0, 0,-1,-1,-1,   0, 0,-1, 0,-1,-1 };
\end{verbatim}
These statements specify 12 neighbours, out of a total of
24 for a site on the four-dimensional fcc lattice.
The first six of these neighbours correspond with
the first line in Eq.~(\ref{eq_nb_fcc}),
the last six with the second line.
The above neighbours are only the ones that have been visited before,
because the loop over sites is performed with the loop over
`{\tt ia}' as the outer-most loop,
and the loop over `{\tt ix}' as the inner-most loop.
For each of the neighbours, {\tt ixp} gives the x-displacement
with respect to the current site.

For non-Bravais lattices, like the diamond lattice,
the {\tt sites} array was given an extra dimension~$l$,
where $l$ runs over the sites in the basis of the lattice.
For the diamond lattice in $d$ dimensions, the basis consists
of 2~sites.
The loop over $l$ was implemented as the inner-most loop.
The neighbours of a site can then be specified as follows.
The sites of type~A are treated differently from the sites
of type~B.
\begin{verbatim}
   int ilp[] = {  1, 1, 1, 1,  -1 };
   int ixp[] = { -1, 0, 0, 0,   0 };
   int iyp[] = {  0,-1, 0, 0,   0 };
   int izp[] = {  0, 0,-1, 0,   0 };
   int iap[] = {  0, 0, 0,-1,   0 };
\end{verbatim}
The first four neighbours are neighbours of the site of type~A.
They are sites of type B, located at a displacement of
$-{\bf x}$, respectively $-{\bf y}$,~$\ldots$
The last neighbour is a neighbour of the site of type B.
It is the other site, type A, in the same basis.


\section{Results}
\label{sec_results}
\noindent
The percolation thresholds of fcc, bcc, and diamond lattices
are given in Tables~\ref{tab_fcc}--\ref{tab_bond},
for several lattice sizes.
All the results were obtained with runs on a
Sun Sparc workstation with 320~Mb internal memory.
The values in Table~\ref{tab_bond} were calculated
by use of the general purpose program.
This program allowed only limited lattice sizes,
for which it is not sure that the scaling relation
\begin{equation}
   | p_c(L) - p_c(\infty) | \propto L^{-1/\nu} 
   \label{eq_scaling}
\end{equation}
holds.
Therefore an estimate was made of a possible systematic error.
This was done by comparing a fit of the scaling relation
to the last three data points
and a fit to the last two data points.
The estimate is given as the error margin
for $p_c(\infty)$ in the table.
During the fitting procedure the value of the exponent~$\nu$
was fixed at $0.88$ in three dimensions,
and $\nu=0.68$, $0.57$, $0.5$ in
$4$, $5$ and $6$ dimensions respectively\cite{stauffer}.

As a check on the results for $p_c(\infty)$, we also
used an alternative method to determine the value for
the percolation thresholds at $L=\infty$.
When we define
\begin{equation}
   \Delta(L) = \sqrt{ | <p_c(L)^2> - <p_c(L)>^2 | } ,
   \label{eq_delta}
\end{equation}
we can use the relation\cite{stauffer}
\begin{equation}
   | <p_c(L)> - p_c(\infty) | \propto \Delta(L) .
\end{equation}
With the use of this relation, the value of $p_c(\infty)$
can be fitted, without knowledge of the value of
the exponent~$\nu$.
The results of this fit are shown in the row marked `check'
in Table~\ref{tab_bond}.
They confirm the values of the row marked `$p_c(\infty)$'.

\setlength{\unitlength}{1cm}
\begin{figure}[tb]
    \epsffile{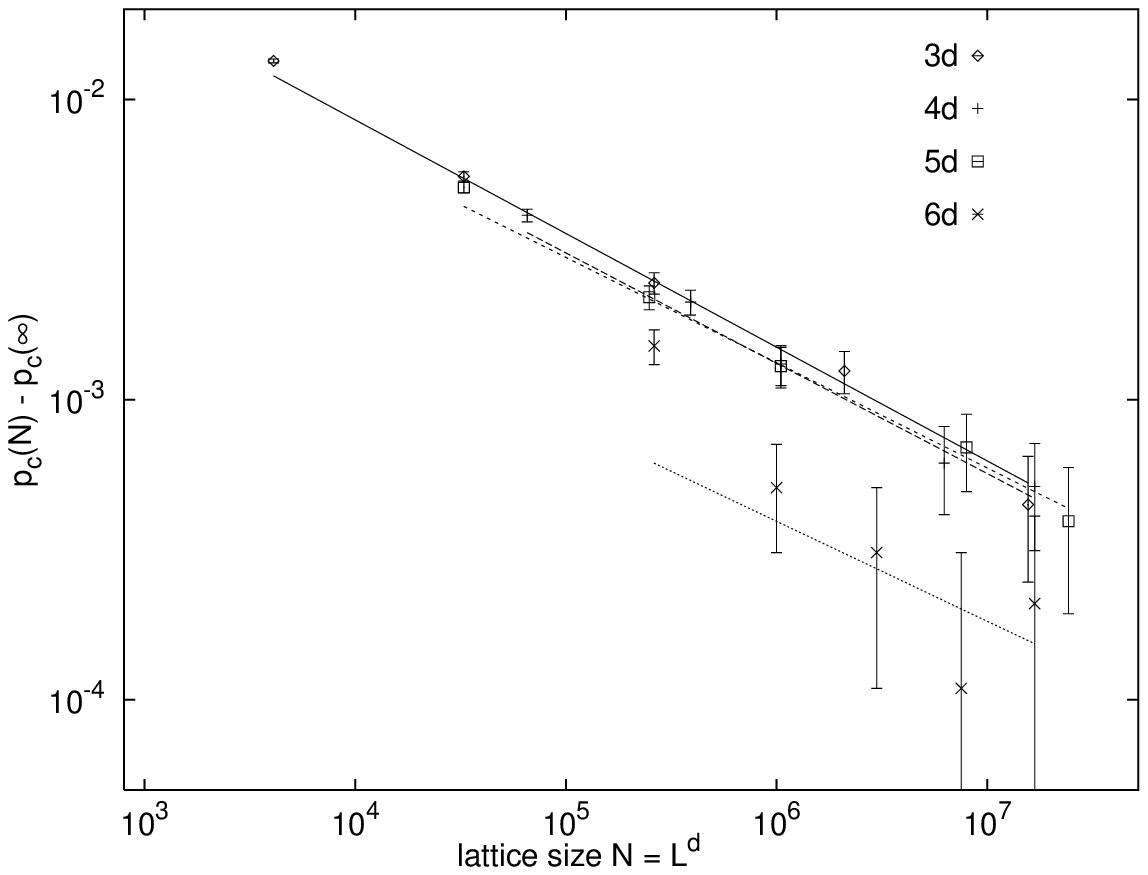}
    \fcaption{\small The scaling of the bcc site percolation
              thresholds as a function of the lattice size
              see Eq.~(\ref{eq_scaling}).
              The fits to the data have a slightly different
              slope, because the exponent~$\nu$ is different
              for each dimension up to $d=6$.
             }
    \label{fig_scaling}
\end{figure}
The results in Tables~\ref{tab_fcc}--\ref{tab_dia} were
obtained using the special purpose program,
which was designed to suit the particular lattice at hand.
Fits to the data points show that the values in the tables are
within the scaling regime, see Fig.~\ref{fig_scaling}.
Therefore the error margins quoted here for $p_c(\infty)$
are the ones indicated by the fit.
The values for the fcc lattices in Table~\ref{tab_fcc}
are different from those reported by Zallen\cite{zallen}
(0.098 in $d=4$ and 0.054 in $d=5$).
However, the cross-check with the general purpose program
confirmed the numbers of Tables~\ref{tab_fcc}--\ref{tab_dia}.
Furthermore, it is not clear
how the Zallen results have been calculated,
nor how large their estimated error margin is.
\begin{figure}[tb]
    \epsffile{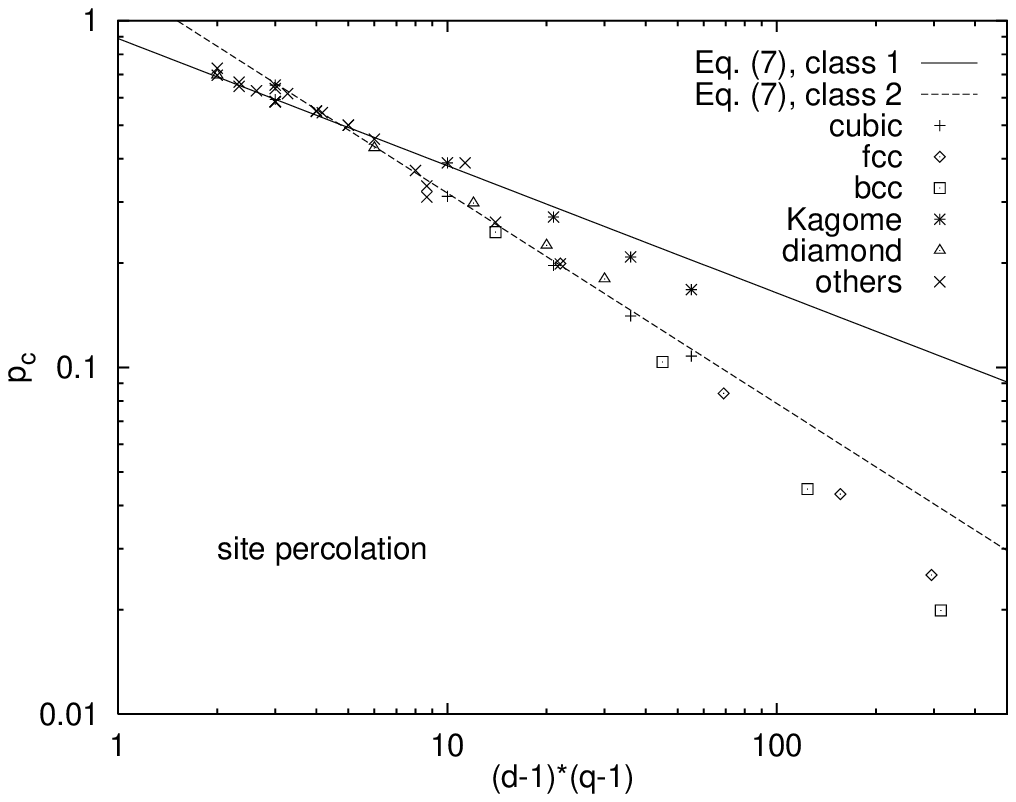}
    \fcaption{\small Site percolation thresholds as a function
              of $(d-1)(q-1)$, see Eq.~(\ref{eq_site})
             }
    \label{fig_site}
\end{figure}
\begin{figure}[tb]
    \epsffile{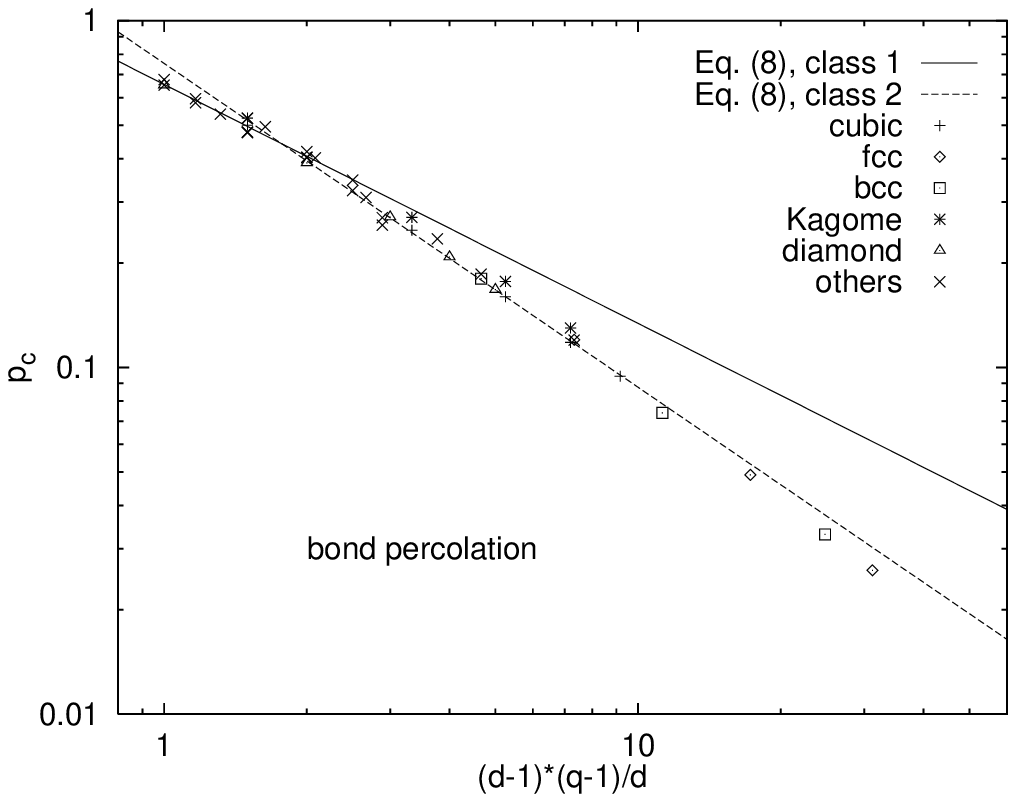}
    \fcaption{\small Bond percolation thresholds as a function
              of $(d-1)(q-1)/d$, see Eq.~(\ref{eq_bond})
             }
    \label{fig_bond}
\end{figure}

\section{Discussion}
\label{sec_discuss}
\noindent
Percolation thresholds of many lattices are listed
in Table~\ref{tab_big}.
\begin{table}[tb]
{\footnotesize
    \begin{center}
    \tcaption{Site ($p_{cs}$) and bond ($p_{cb}$) percolation thresholds
              for various lattices and their dual lattices.
              The deviation of the quantity
              $C = \delta p_{cs}^{-1/a_s}p_{cb}^{1/a_b}/d$
              from 1 is listed under $C-1$.
              Between brackets are error estimates concerning the last digit.
              The thresholds without error margin are exact.
              For the site percolation thresholds on cubic lattices
              in higher dimensions, see also Refs.~\cite{gaunt_76,kirk_76}.
              The lattices in the upper part of the table are two-dimensional,
              including a number of aperiodic lattices;
              the ones in the middle are three-dimensional, and in the lower
              part of the table the higher dimensional lattices are listed.
             }
    \label{tab_big}
    \begin{tabular}{rllrllr}
       \noalign{\smallskip}
       \hline\noalign{\smallskip}
            lattice & $p_{cs}$      & $p_{cb}$      & $C-1$ &
                      $p_{cs}$ dual & $p_{cb}$ dual & $C-1$  \\
       \noalign{\smallskip}\hline\noalign{\smallskip}
       square    & 0.592\,746\,0(5)\cite{ziff_86}
                 & 0.5 \cite{kesten}                            &   0.04 \\
       Kagom\'e  & 0.652\,703\,$\ldots$\cite{sykes}
                 & 0.524\,405\,3(3)\cite{ziff_97}               &$-$0.02
                 & 0.5848(2)\cite{marck_97a}                     
                 & 0.475\,594\,7(3)\cite{ziff_97}               &   0.00 \\
       pentagonal& 0.6471(6)\cite{marck_97a}
                 & 0.5800(6)\cite{marck_97a}                    &   0.01
                 & 0.5502(8)\cite{marck_97a}
                 & 0.4196(6)\cite{marck_97a}                    &$-$0.01 \\
       bowtie    & 0.5475(8)\cite{marck_97a}
                 & 0.404\,518\,$\ldots$\cite{wierman}           &$-$0.05
                 & 0.6653(6)\cite{marck_97a}
                 & 0.595\,481\,$\ldots$\cite{wierman}           &$-$0.02 \\
       triangular& 0.5 \cite{sykes,kesten}
                 & 0.347\,296\,$\ldots$\cite{sykes}             &$-$0.03
                 & 0.6971(2) & 0.652\,703\,$\ldots$\cite{sykes} &$-$0.02 \\
       octagonal & 0.5 \cite{sykes} & 0.3237(6)\cite{marck_97a} &$-$0.12
                 & 0.7297(4)\cite{marck_97a}
                 & 0.6771(6)\cite{marck_97a}                    &$-$0.09 \\
       \noalign{\smallskip}
       Penrose   & 0.5837(3)\cite{yonez}&0.4770(1)\cite{yonez} &   0.01
                 & 0.6381(3)\cite{yonez}&0.5233(2)\cite{yonez} &   0.01 \\
       octag-chem& 0.585(1)\cite{babal1}&0.478(3)\cite{babal2} &   0.01 &&\\
       octag-ferr& 0.543(2)\cite{babal1}&0.402(5)\cite{babal2} &$-$0.04 &&\\
       dodec-chem& 0.628(2)\cite{babal1}&0.538(1)\cite{babal2} &$-$0.01 &&\\
       dodec-ferr& 0.617(3)\cite{babal1}&0.4950(5)\cite{babal2}&   0.01 &&\\
       \noalign{\smallskip}
       cubic       & 0.311\,604(6)\cite{grassberger}
                   & 0.248\,812\,6(5)\cite{ziff_98}        &$-$0.01 \\
       3d-Kagom\'e & 0.3895(2)\cite{marck_98} 
                   & 0.2709(6)\cite{marck_97b}             &$-$0.25 && \\
       diamond     & 0.4301(2)
                   & 0.3893(2)\cite{marck_97a}             &$-$0.06
                   & 0.3895(2)\cite{marck_98}
                   & 0.2350(5)\cite{marck_97b}             &$-$0.35 \\
       trian. stack& 0.2623(2)\cite{marck_97a}
                   & 0.1859(2)\cite{marck_97a}             &$-$0.03
                   & 0.3701(2)\cite{marck_97a}
                   & 0.3093(2)\cite{marck_97a}             &$-$0.05 \\
       bcc         & 0.2458(2)\cite{marck_97a}
                   & 0.180\,287\,5(10)\cite{ziff_98}       &   0.04
                   & 0.4560(6)\cite{marck_97a}
                   & 0.4031(6)\cite{marck_97a}             &$-$0.11 \\
       fcc         & 0.1994(2)\cite{marck_97a}
                   & 0.120\,163\,5(10)\cite{ziff_98}       &$-$0.05
                   & 0.3341(5)\cite{marck_97b}
                   & 0.2703(3)\cite{marck_97b}             &$-$0.03 \\
       hcp         & 0.1990(2)\cite{marck_97a}
                   & 0.1199(2)\cite{marck_97a}             &$-$0.05
                   & 0.3101(5)\cite{marck_97b}
                   & 0.2573(3)\cite{marck_97b}             &   0.04 \\
       \noalign{\smallskip}
       4d-cubic    & 0.196901(5)\cite{balles}
                   & 0.1600(2)\cite{gaunt_78}             &$-$0.01 &&& \\
       4d-Kagom\'e & 0.2715(3)\cite{marck_98} & 0.177(1)  &$-$0.35 &&& \\
       4d-diamond  & 0.2978(2) & 0.2715(3)\cite{marck_98} &$-$0.12 &&& \\
       fcc (`$D_4$') & 0.0842(3) & 0.049(1)               &   0.13 &&& \\
       4d-bcc        & 0.1037(3) & 0.074(1)               &   0.25 &&& \\
       \noalign{\smallskip}

       5d-cubic    & 0.1407(3)\cite{marck_98}
                   & 0.1181(2)\cite{gaunt_78}             &$-$0.04 &&& \\
       5d-Kagom\'e & 0.2084(4)\cite{marck_98} & 0.130(2)  &$-$0.42 &&& \\
       5d-diamond  & 0.2252(3) & 0.2084(4)\cite{marck_98} &$-$0.16 &&& \\
       fcc (`$D_5$') & 0.0431(3) & 0.026(2)               &   0.39 &&& \\
       5d-bcc        & 0.0446(4) & 0.033(1)               &   0.69 &&& \\
       \noalign{\smallskip}

       6d-cubic    & 0.1079(5)\cite{marck_98}
                   & 0.0943(2)\cite{gaunt_78}             &   0.01 &&& \\
       6d-Kagom\'e & 0.1677(7)\cite{marck_98}             &       &&&& \\
       6d-diamond  & 0.1799(5) & 0.1677(7)\cite{marck_98} &$-$0.19 &&& \\
       fcc (`$D_6$') & 0.0252(5) & & &&& \\
       6d-bcc        & 0.0199(5) & & &&& \\
       \noalign{\smallskip}\hline

    \end{tabular}
    \end{center}
}
\end{table}
The results for the deviation of
$C = \frac{\delta}{d} p_{cs}^{-1/a_s}p_{cb}^{1/a_b}$
from 1 is also listed.
When we focus on the results for the invariant~$C$,
it is interesting to compare
the triangular lattice with the octagonal lattice.
Both lattices are isotropic lattices,
with (average) coordination number~$6$.
Both lattices are fully triangulated, and hence have equal
site percolation thresholds\cite{sykes} $p_{cs}=\frac{1}{2}$.
Nevertheless, their bond percolation threshold differ:
$p_{cb} = 2 \sin(\pi/18) = 0.347\,296\,\ldots$\cite{sykes} for
the triangular lattice vs.
$p_{cb} = 0.3237 \pm 0.0006$\cite{marck_97a} for
the octagonal lattice.
Therefore, the value of $C$ for these lattices is
quite different, namely
$C=0.97$ for the triangular and
$C=0.88$ for the octagonal lattice.
Also for the dual of the octagonal lattice,
the value for $C$ deviates substantially
from unity:~$C=0.91$.

One can also compare the cubic lattice and the
3-dim\-ensional Kagom\'e lattice.
The latter has a much higher site percolation threshold
($0.3895 \pm 0.0002$\cite{marck_98} vs.
$0.311\,604 \pm 0.000\,0006$\cite{grassberger}),
but the bond percolation thresholds lie closer together
($0.2706 \pm 0.0009$\cite{marck_97b} vs.
$0.248\,812\,6 \pm 0.000\,000\,5$\cite{ziff_98}).
Although both lattices are isotropic and have
coordination number~$6$,
the respective values for $C$ are $0.99$ and~$0.75$.

The dual of the diamond lattice and the 3-dimensional
Kagom\'e lattice form a special pair too.
It can be shown that the site percolation thresholds
of these lattices are equal, although the lattices
have a different coordination number\cite{marck_97b}.
On the other hand, their bond percolation thresholds
are different, which is reflected in
the values for~$C$ being $0.75$ and $0.65$.
Note further that the dual of the bcc lattice,
which was not incorporated in the
Galam and Mauger study\cite{galam},
also shows a substantial deviation, with $C=0.89$.

Moreover, in high dimensions the dependence of
percolation thresholds on the dimension~$d$ is not universal.
For the proposed invariant, Galam and Mauger assumed
that all thresholds have equal scaling behaviour,
such as the relation $p_c \propto 1/(2d-1)$ 
that was established some time ago
for hypercubic lattices\cite{gaunt_76}.
However, it has been shown recently\cite{marck_98} that
there are $d$-dimensional Kagom\'e lattices
with a different scaling behaviour, namely $p_c \propto 1/d$.
For these lattices the values for $C$ decrease with dimension,
until $C=0.58$ for $d=5$.
Also for the fcc, bcc, and diamond lattices, the behaviour
as a function of dimension has not yet been captured fully.
For each of the lattices the $C$-value deviates more strongly
from unity, as the dimensionality increases.
The scaling of percolation thresholds as a function of
dimension was fitted by Galam and Mauger\cite{galam} to
\begin{equation}
  p_{cs} \propto \left\{ (d-1)(q-1) \right\}^{-a_s}
  \label{eq_site}
\end{equation}
for site percolation thresholds, and
\begin{equation}
  p_{cb} \propto \left\{ \frac{(d-1)(q-1)}{d} \right\}^{-a_b}
  \label{eq_bond}
\end{equation}
for bond percolation thresholds.
The thresholds have been plotted in Figs.~\ref{fig_site}
and~\ref{fig_bond}, together with the Galam and Mauger fit.
The bond percolation thresholds, Fig.~\ref{fig_bond},
all seem to have the same
trend, and the fit captures that trend quite well.
The site percolation thresholds, however, show a much more
complicated behaviour (Fig.~\ref{fig_site}).
The deviations from the Galam and Mauger fit are increasing with
dimension, but also it looks as if there is a spread in the
thresholds, even though they follow a general trend.
The Kagom\'e lattices should be considered separately, since
their scaling behaviour has been shown to differ from the
scaling for hypercubic lattices\cite{marck_98}.

In conclusion, site and bond percolation threshold have
been calculated for several lattices in up to six dimensions.
The scaling behaviour of these percolation thresholds as a
function of dimension is sometimes
different than for other known lattices.
It may therefore be necessary to introduce
more separate classes to allow for the variation in behaviour,
e.g. a separate class for each dimension.
If the deviations would then become smaller again,
the invariant proposed by Galam and Mauger
would still have the remarkable property
of being independent of the coordination number.

\nonumsection{Acknowledgement}
\noindent
The author would like to thank Bob Ziff for his help in compiling
the table of percolation thresholds and the corresponding
references.

\nonumsection{References}


\end{document}